\documentclass[prb,twocolumn,amsmath,showpacs]{revtex4}

\usepackage{graphicx}

\begin{document}
\input epsf

\title {Comment on "Enhanced two-dimensional properties of the four-layered cuprate high-$T_c$ superconductor TlBa$_2$Ca$_3$Cu$_4$O$_y$"}

\author {I. L. Landau$^{1,2}$ H. R. Ott$^{1}$}
\affiliation{$^{1}$Laboratorium f\"ur Festk\"orperphysik, ETH 
H\"onggerberg, CH-8093 Z\"urich, Switzerland}
\affiliation{$^{2}$Kapitza Institute for Physical Problems, 117334 
Moscow, Russia}

\date{\today}

\begin{abstract}

We reanalyze published magnetization  data and demonstrate that the conclusion of the original authors, claiming enhanced two-dimensional properties of the cuprate superconductor TlBa$_2$Ca$_3$Cu$_4$O$_y$, is not supported by the experimental results. Our analysis shows that the magnetic field dependence of the mixed-state magnetization for this particular sample is amazingly close to the results of numerical calculations by E. H. Brandt for an ideal vortex lattice without fluctuations. This good agreement between experiment and theory allows for the evaluation of the absolute values of the upper critical field $H_{c2}(T)$. 

\end{abstract}

\maketitle

We consider the recent study of the magnetization in the superconducting state of  TlBa$_2$Ca$_3$Cu$_4$O$_y$.\cite{kimkim} While we appreciate the high quality of the experimental data, we question most of the conclusions that were made in Ref. \onlinecite{kimkim}. As we argue below, the questionable conclusions are not due to some obvious errors in the data analysis but rather a consequence of the inadequacy of the methods that are traditionally used for the analyses of equilibrium magnetization data obtained for the mixed state of type-II superconductors. 

The tool most often used for such analyses is the Hao-Clem model\cite{h-cl} which represents an analytical approximation to the Ginzburg-Landau theory of the Abrikosov vortex lattice.\cite{abr} It is commonly accepted that this model provides a quantitative description of the mixed state magnetization in high-$T_c$ superconductors (HTSC) and offers the possibility to calculate the thermodynamic critical field $H_c$ and the Ginzburg-Landau parameter $\kappa$ from experimental magnetization data. It has been noticed that the use of this model for the interpretation of  corresponding $M(H)$ data  practically always leads to unphysical results, such as a strong increase of the calculated $\kappa$ values with increasing temperature.\cite{kimkim,hc1,hc2,hc3,hc4,hc5,hc6,hc6a,hc7,hc8,hc9,hc9a,hc10,hc11,hc12,kim}  Instead of rising some doubts in the validity of the model, this feature was taken as evidence of a particularly strong influence of thermal fluctuations on the sample magnetization at temperatures still well below $T_c$.\cite{kimkim,hc1,hc2,hc3,hc4,hc5,hc6,hc6a,hc7,hc8,hc9,hc9a,hc10,hc11,hc12,kim,kogan,kogan2}  However, we recently demonstrated that the fluctuation induced corrections to the magnetization remain negligibly small up to temperatures very close to $T_c$,\cite{lo-hcl} thus suggesting a more general inadequacy of the model. While on previous occasions we could only speculate  on the origin of the failure of the model, we are now able to show that most likely it is the insufficient accuracy in the calculation of $M(H,\kappa)$ using the Hao-Clem model. 

For our data analysis we use a simple scaling procedure developed in Ref. \onlinecite{lo1}. Subsequent work demonstrated that this scaling procedure may successfully be used for analysing the reversible magnetization data, collected for numerous samples of different families of HTSC's and available in the literature.\cite{lo1,lo-c,thomp} One of the main advantages of this scaling approach is that no specific $M(H)$ dependence needs to be assumed a priori. The procedure may thus be used for any mixed-state configuration and any type-II superconductor, independent of the symmetry of the order parameter or the sample geometry. Because of this universality, the scaling procedure does not provide absolute values of $H_{c2}$ but only its relative temperature variation. In order to obtain the $H_{c2}(T)$ curve, it is sufficient, however, to evaluate $H_{c2}$ at one single temperature. In the following, we demonstrate that by comparison of the scaled magnetization curve with results of theoretical calculations by Brandt,\cite{brandt} this last step can indeed be made. 

According to Ref. \onlinecite{lo1}, the relation between the magnetizations at two different temperatures may be written as
\begin{equation}
M(H/h_{c2},T_0) = M(H,T)/h_{c2} + c_0(T)H
\end{equation}
with
\begin{equation}
c_0(T) = \chi_n (T) - \chi_n (T_0),
\end{equation}
where $h_{c2}(T)  = H_{c2}(T)/H_{c2}(T_0)$ and $\chi_n$ is the normal-state magnetic susceptibility. While the first term on the right-hand side of Eq. (1) is universal for any type-II superconductor, the second is introduced to account for the temperature dependent normal-state susceptibility of HTSC's. 

\begin{figure}[t]
 \begin{center}
  \epsfxsize=0.9\columnwidth \epsfbox {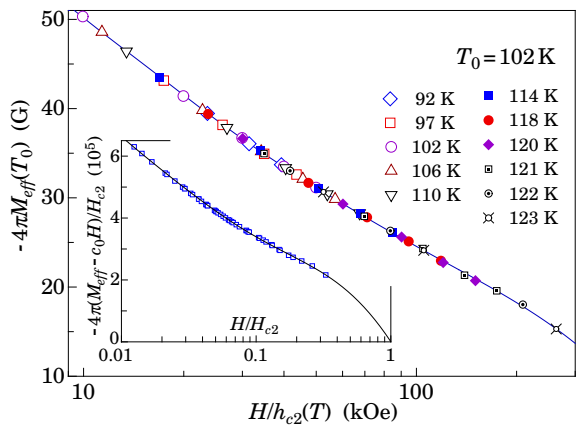}
  \caption{The scaling results for $M(H)$ of a Tl-1234 sample studied in Ref. \onlinecite{kimkim}. The solid line is a theoretical $M(H)$ curve, calculated in Ref. \onlinecite{brandt} and fitted to the data points, as explained in the text. The inset displays the same data, corrected for the normal-state paramagnetic contribution.}
 \end{center}
\end{figure}
\begin{figure}[b]
 \begin{center}
  \epsfxsize=1\columnwidth \epsfbox {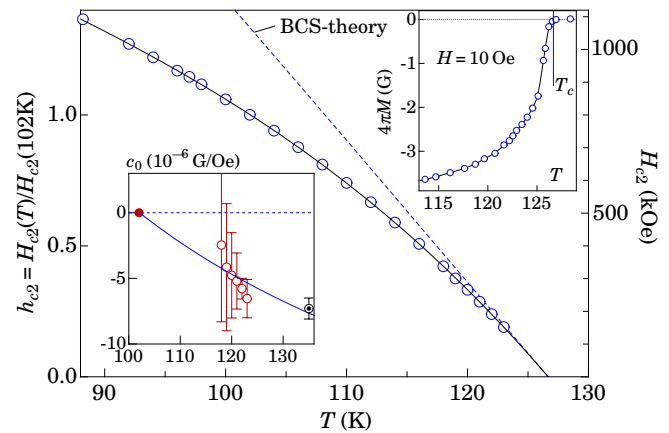}
  \caption{The normalized upper critical field $h_{c2}$ as a function of temperature. The solid line is a guide to the eye. The dashed line represents the $H_{c2}(T)$ curve calculated in the framework of the weak-coupling BCS theory.\cite{gork,wert} The upper inset shows the low-field magnetization data reported in Ref. \onlinecite{kimkim} and connected by a solid line to guide the eye. The vertical line indicates $T_c$, obtained by extrapolating $h_{c2}(T)$ to $h_{c2} = 0$. The lower inset shows the scaling parameter $c_0(T)$. By definition  $c_0(102K) = 0$. The value of $c_0$ at $T = 135$ K was evaluated by fitting the scaled magnetization $M_{eff}(H/h_{c2})$ curve to the result of a theoretical calculation (see text for details). The solid line is the best fit of a Curie law to the data-points.}
 \end{center}
\end{figure}
Fig. 1 shows the magnetization data, scaled according to the mentioned procedure. Here, as in our previous publications, we use $M_{eff}$ to denote the magnetizations calculated from measurements at different temperatures employing Eq. (1). $M_{eff}(H)$ represents the equilibrium magnetization curve for $T = T_0$.\cite{lo1,lo-irr} As may be seen in Fig. 1, all available data perfectly merge onto a single curve with virtually no scatter. No corrections for possible fluctuation effects need to be considered and hence we disagree with the conclusion of Ref. \onlinecite{kimkim}, which claims that fluctuation induced effects on the magnetization are already significant at $T = 115$ K. 

The temperature dependencies of the scaling parameters $h_{c2}$ and $c_0$ are presented in Fig. 2. Extrapolating $h_{c2}(T)$ to $h_{c2} = 0$, we obtain the value of $T_c = 126.7 \pm 0.5$ K. As may be seen in the upper inset of Fig. 2, this value is consistent with the low-field magnetization data. For this particular sample, the temperature dependence of $\chi_n$ turns out to be weak and its contribution to the total magnetization is significant only at $T \ge 118$ K. This is the reason why $c_0$ can be evaluated only in a narrow temperature range between 118 and 123 K and with a rather large uncertainty (see the lower inset in Fig. 2). 

As far as we are aware, the only reliable calculations of the magnetic response of type-II superconductors in the mixed state  were presented in Ref. \onlinecite{brandt}. The results of this work for $5 \le \kappa \le 200$ are shown in Fig. 3 as $\kappa^2M/B_{c2}$ versus $B/B_{c2}$ ($B$ is the magnetic induction). For comparison, also a typical result of Ref. \onlinecite{h-cl} is included in the same figure. As may be seen, for $B/B_{c2} \le 0.2$ the $M(B)$ curve calculated by employing of the Hao-Clem model deviates significantly from those that are obtained by more accurate numerical calculations. This difference and the unjustified lack of considering the temperature dependence of the normal-state magnetic susceptibility  seem to be responsible for the resulting $\kappa(T)$ curves exhibiting a strong increase of $\kappa$ at higher temperatures. Because, as may be seen in the inset of Fig. 1, most of the $M(H)$ data points that are available for the analysis were taken in magnetic fields $H < 0.3H_{c2}$, the disagreement between the two types of calculations needs to be taken seriously. The agreement between the Hao-Clem model and Brandt's calculations in higher fields is, for our purpose, irrelevant.
\begin{figure}[h]
 \begin{center}
  \epsfxsize=0.9\columnwidth \epsfbox {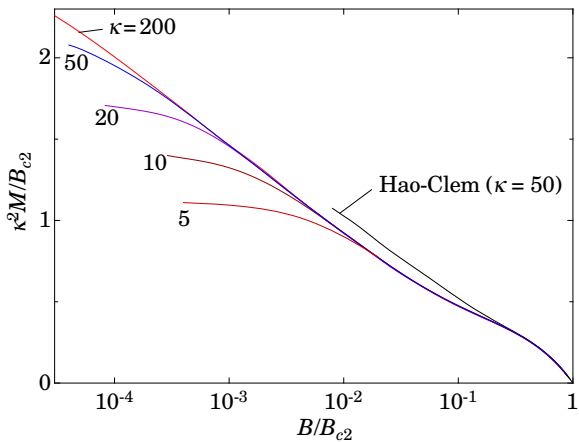}
  \caption{The results of numerical calculations\cite{brandt} of the magnetization of the Abrikosov vortex lattice for several different values of the Ginzburg-Landau parameter $\kappa$ (indicated near the curves).  The analogous curve, calculated in Ref.  \onlinecite{h-cl}, is shown for comparison.}
 \end{center}
\end{figure}

For $\kappa \ge 20$ and $B/B_{c2} \ge 10^{-3}$ all the curves calculated according to Ref. \onlinecite{brandt} merge onto a single $\kappa^2M/B_{c2}$ versus $B/B_{c2}$ curve. Therefore this universal curve may be used for fitting experimental data, using $\kappa$ and $B_{c2}$ as fit-parameters.\cite{B-H} In order to fit the scaled magnetization curve presented in Fig. 1 in this manner, it must be taken into account that this curve, in addition to  the diamagnetic response $M_d$ of the mixed state, also includes a paramagnetic contribution corresponding to $c_0(135K) =\chi_n(135K) - \chi_n (T_0)$.\cite{ftnt} Because from the data presented in the lower inset of Fig. 2, the value of $c_0(135K)$ cannot be evaluated with sufficient accuracy, we introduce it as an additional fit parameter. As may be seen in Fig. 1, the theoretical curve fitted in such a way follows perfectly the data points over the entire covered range of magnetic fields. The fitting procedure leads to the following values of the parameters: $H_{c2}(102K) = (800 \pm 80)$ kOe, $\kappa^{\ast} = (118 \pm 5)$ (we use here $\kappa^{\ast}$ because, as we discuss below, the value estimated in such a way may differ significantly from the true value of the Ginzburg-Landau parameter for this compound), $c_0(135K) = - (7.3 \pm 0.8)$ G/Oe. The indicated error margins reflect only the quality of the fit and do not include experimental and possible systematic errors in the magnetization data presented in Ref. \onlinecite{kimkim}.

Evidently, the quoted fit value for $c_0(135K)$ matches very well the trend of the other $c_0$ values that were obtained via the scaling of the magnetization data (see lower inset in Fig. 1).  This is encouraging for two reasons. First, this consistency together with the very high quality of the fit demonstrates that the comparison of $M_{eff}(H)$ with the results of Brandt's calculations is indeed justified. Second, it provides an additional confirmation of the validity of our scaling procedure. 

Although the agreement between experimental results and theory, demonstrated in Fig. 1, is very good, several important remarks considering the reliability of the numerical values of $H_{c2}(102K)$ and $\kappa^{\ast}$ have to be made. (i) The obtained value of $\kappa^{\ast} = 118$ does not represent the real value of the Ginzburg-Landau parameter $\kappa$ for this compound. Indeed, the calculations in Ref. \onlinecite{brandt} were made assuming a zero demagnetizing factor $n$, which is not the case for the experiment that we consider here. Nonzero values of $n$ reduce the sample magnetization at a given value of $B/B_{c2}$, i.e., $M = 0$ for $n = 1$. As may be seen from Fig 3, for magnetic fields $H \gg H_{c1}$, $M \sim \kappa^{-2}$. In other words, an increase of $n$ from $n = 0$ has exactly the same impact on the sample magnetization as an enhancement of $\kappa$. This is why the true value of $\kappa$ is certainly smaller than $\kappa^{\ast} = 118$. (ii) The evaluation of  $H_{c2}(T_0)$ is achieved by extrapolating the diamagnetic part $M_d(H)$ at $T = T_0$ to $M_d(H) = 0$. Because the range of the experimental data points covers only magnetic fields $H \lesssim 0.3 H_{c2}$ (see inset of Fig. 1), even relatively small experimental errors, which cannot be taken into account in our calculations, may result in a significant error in the absolute value of $H_{c2}(102K)$.

In conclusion, we have shown that the magnetization data presented in Ref. \onlinecite{kimkim} may perfectly well be described by the Ginzburg-Landau theory for an isotropic type-II superconductor. No significant contribution of fluctuation effects on the magnetization data for $T \le 123$ K $\approx 0.97T_c$ may be identified. Therefore, all the statements in Ref. \onlinecite{kimkim} that are a consequence of the analysis of the high-temperature magnetization data, including the consideration of fluctuation induced contributions to $M(H)$, are not supported by the available experimental results. For the same reason, the title of Ref. \onlinecite{kimkim}, claiming a strongly two-dimensional character of  superconductivity in Tl-1234 is not tenable. Our analysis provides the following characteristic parameters of the Tl-1234 sample investigated in Ref. \onlinecite{kimkim}: $T_c = 127 \pm 0.5$ K (instead of 128-130.5 K quoted in Ref. \onlinecite{kimkim}), $(dH_{c2}/dt)_{T=T_c} = (42.5 \pm 4)$ kOe/K (instead of 22.3 kOe/K in Ref. \onlinecite{kimkim}) and $\kappa < 118$, which is, in fact, close to the value quoted in Ref. \onlinecite{kimkim}. Finally,  we note that the use of the weak-coupling BCS theory,\cite{gork,wert} in this context the so called Werthamer-Helfand-Hohenberg formula for calculating the value of $H_{c2}$ for $T = 0$, as it is done in Ref. \onlinecite{kimkim} and many other studies of HTSC compounds, is not at all justified. Indeed, as may clearly be seen in Fig. 2, the real $H_{c2}(T)$ curve deviates substantially from the prediction of the BCS theory.


\begin{thebibliography}{71}

\bibitem{kimkim} K.-H. Kim, H.-J. Kim, S.-I. Lee, A. Iyo, Y. Tanaka, K. Tokiwa, T. Watanabe, Phys. Rev. B {\bf 70}, 92501 (2004).

\bibitem{h-cl} Z. Hao and J. R. Clem, Phys. Rev. Lett. {\bf 67}, 2371(1991). Z. Hao and J. R. Clem, M. W. McElfresh, L.Civale, A. P. Malozemoff, and F. Holtzberg, Phys. Rev. B {\bf 43}, 2844 (1991).

\bibitem{abr} A. A. Abrikosov, Zh. Eksp. Teor. Fiz. {\bf 32}, 1442 (1957) [Sov. Phys. JETP {\bf 5}, 1174 (1957).

\bibitem{hc1} Qiang Li, M. Suenaga, Junho Gohng, D. K. Finnemore, T. Hikata, and K. Sato, Phys. Rev. B {\bf 46}, R3195 (1992).

\bibitem{hc2} J. H. Cho, Zhidong Hao, and D. C. Johnston, Phys. Rev. B {\bf 46}, R8679 (1992).

\bibitem{hc3} Qiang Li, M. Suenaga, T. Kimura, and K.Kishio, Phys. Rev. B {\bf 47}, 2854 (1993).

\bibitem{hc4} Qiang Li, M. Suenaga, T. Kimura,  and K.Kishio, Phys. Rev. B {\bf 47}, 11384 (1993).

\bibitem{hc5} D. N. Zheng, A. M. Campbell, and R. S. Liu, Phys. Rev. B {\bf 48}, 6519 (1993).

\bibitem{hc6} Qiang Li, K. Shibutani, M. Suenaga, I. Shigaki, and R. Ogawa, Physica B {\bf 194-196}, 1501 (1994).

\bibitem{hc6a} Y. C. Kim,  J. R. Thompson, J. G. Ossandon, D. K. Christen, M. Paranthaman, Phys. Rev. B {\bf 51}, 11767 (1995).

\bibitem{hc7} Mung-Seog Kim, Sung-Ik Lee, Seong-Cho Yu, and Nam H. Hur,  Phys. Rev. B {\bf 53}, 9460 (1996).

\bibitem{hc8} J. R. Thompson, J. G. Ossandon, D. K. Christen, M. Paranthaman, E. D. Specht, and Y. C. Kim, Phys. Rev. B {\bf 54}, 7505 (1996).

\bibitem{hc9} Yi Zhuo, Jae-Hyuk Choi, Mung-Seog Kim, Wan-Seon Kim, Z. S. Lim, Sung-Ik Lee, Sergey Lee, Phys. Rev. B {\bf 55}, 12719 (1997).

\bibitem{hc9a} Yi Zhuo, Jae-Hyuk Choi, Mung-Seog Kim, Jin-Nam Park,  Myong-Kwang Bae, Sung-Ik Lee, Phys. Rev. B {\bf 56}, 8381 (1997).

\bibitem{hc10} Mung-Seog Kim, Sung-Ik Lee, Seong-Cho Yu, I. Kuzemskaya, E. S. Itskevich, and K. A. Lokshin,  Phys. Rev. B {\bf 57}, 6121 (1996).

\bibitem{hc11} M. Y. Cheon, G. C. Kim, B. J. Kim, and Y. C. Kim, Physica C {\bf 302}, 215 (1998).

\bibitem{hc12} Yi Zhuo, Su-Mi Oh, Jae-Hyuk Choi, Mun-Seog Kim, Sung-Ik Lee, N. P. Kiryakov, M. S. Kuznetsov, and Sergey Lee, Phys. Rev. B {\bf 60}, 13094 (1999).

\bibitem{kim} Heon-Jung Kim, P. Chowdhury, In-Sun Jo, and Sung-Ik Lee, Phys. Rev. B {\bf 66}, 134508 (2002).

\bibitem{kogan} V. G. Kogan, M. Ledvij, A. Yu. Simonov, J. H. Cho, D. C. Johnston, Phys. Rev. Lett. {\bf 70}, 1870 (1993).

\bibitem{kogan2} V. G. Kogan, A. Gurevich, J. H. Cho, D. C. Johnston, M. Xu, J. R. Thompson, A. Martynovich, Phys. Rev. B {\bf 54}, 12386 (1996).


\bibitem{lo-hcl} I. L. Landau and H. R. Ott, Physica C {\bf 411}, 83 (2004).

\bibitem{lo1} I. L. Landau and H. R. Ott, Phys Rev. B {\bf 66}, 144506 (2002).

\bibitem{lo-c} I. L. Landau and H. R. Ott, Physica C {\bf 385}, 544 (2003).

\bibitem{thomp} J. R. Thompson, J. G. Ossandon, L. Krusin-Elbaum, D. K. Christen, H. J. Kim, K. J. Song, K. D. Sorge, and J. L. Ullmann, Phys. Rev. B {\bf 69}, 104520 (2004).

\bibitem{lo-irr} I. L. Landau and H. R. Ott, Phys Rev. B {\bf 67}, 92505 (2003).

\bibitem{brandt} E. H. Brandt, Phys. Rev. Lett. {\bf 78}, 2208 (1997).  E. H. Brandt, Phys. Rev. B {\bf 68}, 054506 (2003).

\bibitem{B-H} For the experimental data in Ref. \onlinecite{kimkim}, the difference $(H - B)$ is negligibly small. This is why we use $H$ in our presentation of the experimental data. 

\bibitem{ftnt} From the presentation of the original data (Fig. 2 of Ref. \onlinecite{kimkim}) it is clear that the paramagnetic contribution to $M$ at $T = 135$ K was subtracted from the experimental data. Because the normal-state magnetic susceptibility is temperature dependent, the corresponding corrections to the mixed state magnetization ought to be taken into account at lower temperatures.

\bibitem{gork} L. P. Gor'kov, Zh. Eksp. Teor. Fiz. {\bf 37}, 833 (1959) [Soviet Phys.ÐJETP {\bf 10}, 593 (1960)].

\bibitem{wert} E. Helfand and N. R. Werthamer, Phys Rev. {\bf 147}, 288 (1966), N. R. Werthamer, E. Helfand and G. Hohenberg, {\it ibid.} {\bf 147}, 295 (1966).


\end{thebibliography}
\end{document}